\documentclass[twoside,english,5p,times]{elsarticle}
\PassOptionsToPackage{vlined,ruled}{algorithm2e}
\usepackage[T1]{fontenc}
\usepackage[latin9]{inputenc}
\usepackage{babel}
\usepackage{prettyref}
\usepackage{url}
\usepackage{algorithm2e}
\usepackage{graphicx}
\usepackage[unicode=true]
 {hyperref}

\makeatletter
\journal{Parallel Computing}

\usepackage{hyperref}
\usepackage[switch]{lineno}

\newrefformat{chap}{\hyperref[#1]{Chapter~\ref*{#1}}}
\newrefformat{sec}{\hyperref[#1]{Section~\ref*{#1}}}
\newrefformat{subsec}{\hyperref[#1]{Section~\ref*{#1}}}
\newrefformat{app}{\hyperref[#1]{\ref*{#1}}} % ref adds Appendix
\newrefformat{lst}{\hyperref[#1]{Listing~\ref*{#1}}}
\newrefformat{fig}{\hyperref[#1]{Figure~\ref*{#1}}}
\newrefformat{tab}{\hyperref[#1]{Table~\ref*{#1}}}
\newrefformat{eqn}{\hyperref[#1]{Equation~\ref*{#1}}}
\hypersetup{colorlinks=true, breaklinks=true, linkcolor=darkblue, urlcolor=darkblue, citecolor=darkblue}

\def\ps@pprintTitle{%
 \let\@oddhead\@empty
 \let\@evenhead\@empty
 \def\@oddfoot{}%
 \let\@evenfoot\@oddfoot}

\makeatother

\begin{document}
\global\long\def\B{\:\mathrm{B}}%
\global\long\def\kB{\:\mathrm{kB}}%
\global\long\def\MB{\:\mathrm{MB}}%
\global\long\def\GB{\:\mathrm{GB}}%
\global\long\def\TB{\:\mathrm{TB}}%
\global\long\def\PB{\:\mathrm{PB}}%
\global\long\def\MiB{\:\mathrm{MiB}}%
\global\long\def\GiB{\:\mathrm{GiB}}%
\global\long\def\TiB{\:\mathrm{TiB}}%
\global\long\def\PiB{\:\mathrm{PiB}}%
\global\long\def\GBps{\:\mathrm{GBps}}%
\global\long\def\bit{\:\mathrm{bit}}%
\global\long\def\bits{\:\mathrm{bits}}%
\global\long\def\GHz{\:\mathrm{GHz}}%
\global\long\def\Hz{\:\mathrm{Hz}}%
\global\long\def\rate{\nu}%
\global\long\def\indegree{K}%
\global\long\def\vps{MT}%
\global\long\def\numax{\nu_{\text{max}}}%
\global\long\def\NE{N_{\text{E}}}%
\global\long\def\NI{N_{\text{I}}}%
\global\long\def\E{\text{E}}%
\global\long\def\I{\text{I}}%
\global\long\def\pF{\mathrm{\:pF}}%
\global\long\def\mV{\mathrm{\:mV}}%
\global\long\def\pA{\mathrm{\:pA}}%
\global\long\def\s{\:\mathrm{s}}%
\global\long\def\ms{\:\mathrm{ms}}%
\global\long\def\us{\:\mu\text{sec}}%
\global\long\def\FLOPS{\:\mathrm{FLOPS}}%
\global\long\def\PFLOPS{\:\mathrm{PFLOPS}}%
\global\long\def\TFLOPS{\:\mathrm{TFLOPS}}%
\global\long\def\percent{\:\%}%
\global\long\def\NSEA{N_{\text{SEA}}}%
\global\long\def\NSWP{N_{\text{SWP}}}%

\begin{frontmatter}{}

\title{Routing brain traffic through the von Neumann bottleneck: Efficient
cache usage in spiking neural network simulation code on general purpose
computers}

\author[inm6,rwth]{J.~Pronold\corref{cor1}}

\ead{j.pronold@fz-juelich.de}

\author[pylunibe]{J.~Jordan}

\author[jsc]{B.J.N.~Wylie}

\author[rccs]{I.~Kitayama}

\author[inm6,physrwth,psychrwth]{M.~Diesmann}

\author[nmbu]{S.~Kunkel\corref{cor2}}

\ead{susanne.kunkel@nmbu.no}

\cortext[cor1]{Corresponding author}

\cortext[cor2]{Principal corresponding author}

\address[inm6]{Institute of Neuroscience and Medicine (INM-6) and Institute for
Advanced Simulation (IAS-6) and JARA Institute Brain Structure-Function
Relationships (INM-10), Jülich Research Centre, Jülich, Germany}

\address[rwth]{RWTH Aachen University, Aachen, Germany}

\address[pylunibe]{Department of Physiology, University of Bern, Bern, Switzerland}

\address[jsc]{Jülich Supercomputing Centre, Jülich Research Centre, Jülich, Germany}

\address[rccs]{RIKEN Center for Computational Science, Kobe, Japan}

\address[physrwth]{Department of Physics, Faculty 1, RWTH Aachen University, Aachen,
Germany}

\address[psychrwth]{Department of Psychiatry, Psychotherapy and Psychosomatics, Medical
Faculty, RWTH Aachen University, Aachen, Germany}

\address[nmbu]{Faculty of Science and Technology, Norwegian University of Life Sciences,
Ås, Norway}
\begin{abstract}
Simulation is a third pillar next to experiment and theory in the
study of complex dynamic systems such as biological neural networks.
Contemporary brain-scale networks correspond to directed graphs of
a few million nodes, each with an in-degree and out-degree of several
thousands of edges, where nodes and edges correspond to the fundamental
biological units, neurons and synapses, respectively. When considering
a random graph, each node's edges are distributed across thousands
of parallel processes. The activity in neuronal networks is also sparse.
Each neuron occasionally transmits a brief signal, called spike, via
its outgoing synapses to the corresponding target neurons. This spatial
and temporal sparsity represents an inherent bottleneck for simulations
on conventional computers: Fundamentally irregular memory-access patterns
cause poor cache utilization. Using an established neuronal network
simulation code as a reference implementation, we investigate how
common techniques to recover cache performance such as software-induced
prefetching and software pipelining can benefit a real-world application.
The algorithmic changes reduce simulation time by up to $50\%$. The
study exemplifies that many-core systems assigned with an intrinsically
parallel computational problem can overcome the von Neumann bottleneck
of conventional computer architectures.
\end{abstract}
\begin{keyword}
spiking neural networks \sep large-scale simulation \sep cache performance
\sep distributed computing \sep parallel computing \sep memory
access bottleneck
\end{keyword}

\end{frontmatter}{}

\section{Introduction}

\label{sec:intro}

Irregular access to large amounts of memory challenges the von Neumann
computer architecture. Distributed applications typically make use
of systems with hybrid parallelization, using message passing libraries
for the communication between compute nodes and multi-threading to
employ the computational cores in each node. In this contribution,
we investigate as an extreme real-world example application simulation
code for biological neural networks.

Such networks correspond to graphs. The graph representing the neurons
as nodes and their contact points, called synapses, as directed edges
is sparse and complex. In the mammalian brain a neuron establishes
several thousands or even more than ten-thousand of incoming synapses
and the number of outgoing synapses is of the same order. This corresponds
to a graph, where both in-degree and out-degree of each node are of
the order of $1000$ or even $10,000$.

The brain mantle, called cerebral cortex, contains the neuronal cell
bodies. Tangential to the cortical surface, the probability of two
cortical neurons to establish a contact is approximately $0.1$ within
a distance of one millimeter, but it declines rapidly for longer distances.
Half of a neuron's outgoing connections are not local but target neurons
at distant locations forming a hierarchically organized architecture
\citep[for an example see][]{Schmidt18_1409}. Due to the sheer number
of neurons in the brain, the probability of any pair sharing an edge
is vanishingly small.

The interaction between neurons is mediated through synapses by point-like
events, called spikes. Spike events are sparse in time given that
neurons emit a single or few spikes per second while the time constants
of single-neuron dynamics are in the range of milliseconds and also
behavior is organized by the brain on a sub-second time scale. The
time required to decide whether an image contains a living object
is $180\ms$ for monkeys and $270\ms$ for humans, the interval between
eye movements is about $250\ms$, and humans utter about two words
per second. Neuroscientists hypothesize that sparse, distributed brain
activity supports this systems-level behavior.

Synapses outnumber the neurons by three to four orders of magnitude
and thereby consume a significant part of computer memory in simulations
of spiking neural networks. The strength of the interaction mediated
by a synapse can change over time depending on the activities of the
presynaptic and the postsynaptic neurons and third factors like neuromodulators
and the membrane potential of the postsynaptic neuron. This dynamics,
called synaptic plasticity, is a key mechanism of system-level learning.
For the purpose of this investigation we assume that each synapse
maintains a state variable representing the coupling strength, which
we refer to as synaptic weight, but plastic processes are not considered.
Moreover, synaptic transmission of spikes entails a delay, which is
the time interval between the presynaptic neuron emitting a spike
and the spike taking effect on the postsynaptic neuron. Depending
on the spatial distance between presynaptic and postsynaptic neuron
the delay can be shorter than $0.1\ms$ or longer than $10.0\ms$.
This study considers homogeneous delays of $1.5\ms$.

The spikes emitted by model neurons represent the sharp voltage transients
of biological neurons, called action potentials. Models describing
neuronal networks at the level of resolution of neurons and synapses
represent individual neurons by a small system of differential equations.
Often the system is linear and all non-linearity is condensed in a
threshold operation on the state vector generating the point-like
event. In our case the subthreshold dynamics can be integrated exactly,
limiting the workload in terms of floating point operations.

Over the past two decades simulation tools in computational neuroscience
have increasingly embraced a conceptual separation of generic simulation
engines and specific models of neuronal networks \citep{Einevoll19_735}.
Many different models can thus be simulated with the same simulation
engine. This enables the community to separate the life cycle of a
simulation engine from those of specific models and to maintain and
further develop simulation engines as an infrastructure. Furthermore,
the separation facilitates the cross-validation of simulation engines.

Before the dynamical state of a model of a neuronal network can be
propagated an instance of the model needs to be created in computer
memory. Often network models are concisely defined by probabilistic
construction rules rather than explicit adjacency lists. Therefore,
in simulations of neuronal networks we distinguish between the phase
of network construction and the actual simulation phase, where state
propagation takes place. The former is a research topic on its own
\citep{Ippen2017_30}. The present work concentrates on the simulation
phase. While network construction may take relevant amounts of wall-clock
time, the simulation time scales with the biological time span to
be covered by the model, but network construction does not. Propagating
the dynamical state of the network in time involves three repeating
phases \citep{Morrison05a,Morrison08_267}. The first, termed update,
advances the state of the neurons by a time interval corresponding
to the minimal synaptic delay in the system, where even smaller update
steps at the level of individual neurons are possible. The second,
communication, is concerned with distributing the spikes that have
occurred in this time interval to the compute nodes and threads hosting
the respective target neurons. The subsequent spike-delivery phase
routes the spikes arriving at a compute node via the representations
of the corresponding synapses to their target neurons. Our investigation
concentrates on this final phase of the cycle.

The combination of irregular spiking activity and sparse connectivity,
leads to a practically random memory-access pattern during spike delivery.
Seemingly this is a worst case situation for the von Neumann architecture
where for any computation the content of a respective memory unit
has to be transported to the central processing unit and the result
needs to be transported back. Other disciplines, such as graph processing
\citep{lumsdaine2007challenges} and main memory database systems
\citep{ailamaki1999dbmss,manegold2000optimizing} suffer from frequent
and unpredictable main memory access as well.

\begin{figure}
\begin{centering}
\includegraphics[width=1\columnwidth]{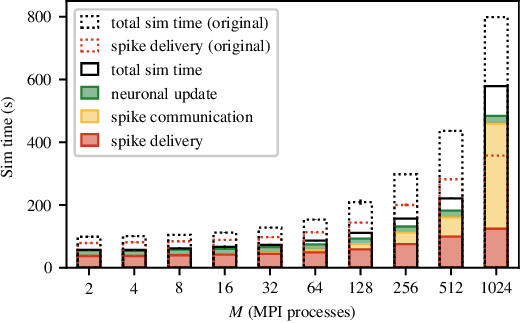}
\par\end{centering}
\caption{Contributions to simulation time (sim time) for spiking neural network
simulations compared to the situation prior to optimization. Weak-scaling
experiment running $2$ MPI processes per compute node and $12$ threads
per MPI process, with a workload of $125,000$ neurons per MPI process
(network model see \prettyref{subsec:network-model}). The network
dynamics is simulated for $1\protect\s$ of biological time; spikes
are communicated across MPI processes every $1.5\protect\ms$. Solid
bars show total sim time (black outline), time spent on spike delivery
(red), communication of spike data (yellow), and neuronal update (green)
after refactoring \citep{pronold2022} and using a combination of
the optimizations of spike delivery discussed in this study (\prettyref{alg:bwRB*},
\prettyref{alg:bwTS}). Dashed bars show total sim time (black dashed)
and time spent on spike delivery (red dashed) of the original code
(\textbf{Algorithm ORI} in \citep{pronold2022}); original communication
and update times (Figure 3 of \citep{pronold2022}) omitted for clarity.
Error bars (visible only at $128$ MPI processes for the original
sim time) indicate standard deviation over three repetitions. Timings
obtained via manual instrumentation of the source code; measured on
JURECA CM (\prettyref{subsec:Systems}).\label{fig:conclusions-simtime-contributions}}
\end{figure}
Let us consider a concrete example for illustration. In weak scaling
of simulations with the same number of neurons per MPI process, spike
delivery dominates simulation time independent of the number of MPI
processes employed (\prettyref{fig:conclusions-simtime-contributions}).
In the regime from $2$ to $512$ MPI processes the time required
for spike delivery almost quadruples (factor of $3.9$). The refactoring
efforts described in a technical companion paper \citep{pronold2022}
and the optimizations discussed in this article reduce the dependence
of spike delivery on the number of MPI processes. Not affected by
these changes to the original spike-delivery algorithm are neuronal
update and communication (cf. Figure 3 of \citep{pronold2022}). Additionally,
the absolute time for neuronal update remains unchanged throughout
as the number of neurons per MPI process is fixed. Beyond $512$ MPI
processes the relative contribution of spike delivery to simulation
time drops below $50\%$ because the time required for communication
increases.

All simulation code analyzed in this study already provides optimizations
reducing the number of spikes that MPI processes need to exchange
in small to medium scale simulations (\prettyref{subsec:reference-deliver-algo};
see Section 3.3 in \citealp{Jordan18_2}). The code exploits the lesser
degree of distribution across processes in this regime where the average
number of outgoing synapses per neuron outnumbers the total number
of MPI processes. This reduces both communication times and spike-delivery
times, which in turn results in shorter overall simulation times (\prettyref{fig:conclusions-simtime-contributions}).
The present work addresses this practically relevant intermediate
regime, where the effect of the optimizations gradually diminishes
with the degree of distribution across processes, which impacts scalability.
Nevertheless, the code scales well on modern supercomputers (see Figure
7C in \citealp{Jordan18_2}; 5g-sort) as the optimizations reach their
limit in the large-scale regime as the outgoing synapses of each neuron
are almost fully distributed across processes. A technical companion
paper \citep{pronold2022} derives an analytical expression for this
transition to a fully distributed network under the constraints of
weak scaling.

\begin{figure}
\begin{centering}
\includegraphics{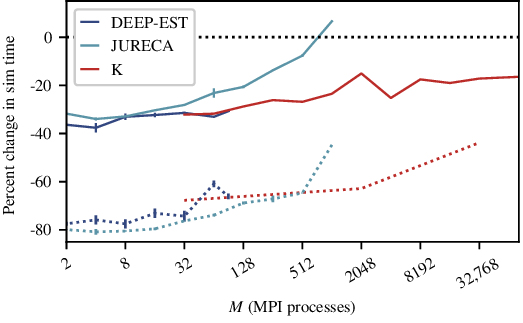}
\par\end{centering}
\caption{Relative change in simulation time after preparatory refactoring of
spike-delivery algorithm (\prettyref{alg:REF}) as a function of the
number of MPI processes $M$. Weak scaling of benchmark network model
(\prettyref{subsec:network-model}) in linear-log representation.
Same configuration as in \prettyref{fig:conclusions-simtime-contributions}
for DEEP-EST CM (blue) and JURECA CM (turquoise) systems; error bars
show standard deviation based on $3$ repetitions. Data of K computer
(red) for $1$ MPI process per compute node, $8$ threads per MPI
process, and $18,000$ neurons per MPI process (systems identified
in \prettyref{subsec:Systems}). Black dotted line at zero indicates
performance of original code (\textbf{ORI} in \citep{pronold2022}).
Dotted curves with corresponding colors indicate hypothetical limit
to the decrease in sim time defined by the contribution of spike-delivery
time to sim time.\label{fig:results-REF}}
\end{figure}
The success of an optimization needs to be evaluated in the light
of potential gain. The maximal gain by an optimization to a specific
part of code is limited by the contribution of this part of code to
the overall runtime. Conversely, if a different part of the code dominates
the runtime, the optimization may go unnoticed. The spike-delivery
algorithm that serves as a reference in this study (\prettyref{subsec:reference-deliver-algo})
already benefits from the refactoring described in a technical companion
paper \citep{pronold2022}. The refactoring significantly reduces
simulation time (\prettyref{fig:results-REF}) showing different effectiveness
on the tested systems (\prettyref{subsec:Systems}). However, over
the full range there is room for further improvement to spike delivery
by at least the same amount.

Recently, \citeauthor{cremonesi2020analytic} made major progress
by deriving a detailed analytic performance model \citep{cremonesi2020analytic,cremonesi2020understanding}.
The model reveals specific algorithmic and hardware bottlenecks for
three classes of neuronal network models mapping out the field of
computational neuroscience. Latency of the communication network and
memory bandwidth are identified as the most severe bottlenecks in
large-scale simulations. The authors find that in models of the type
discussed in the present study the spike-delivery phase is the most
expensive one once the data do not fit into the cache hierarchy anymore.
The most severe hardware constraint is the saturation of memory bandwidth
which is driven by the memory-latency effect of the irregular access
patterns. While the CPU is capable of issuing many memory accesses
in advance to partially hide latency of processing spikes, the authors
expect that two established strategies, pipelining and prefetching,
cannot effectively hide the latency introduced by the non-contiguous
data accesses.

Prefetching attempts to hide cache misses and memory stalls \citep{mittal2016survey}
by overlapping memory access with computation. In general, two types
can be distinguished: hardware-induced and software-induced prefetching.
The former relies on the underlying hardware to detect patterns in
memory access such that the hardware can take care of getting the
data into the cache just when it is needed. The latter relies on hints
in the source code via prefetch instructions indicating what data
should be loaded into cache. This is a promising technique when hardware
predictions fail but the access pattern can still be known by the
developer a priori.

In the past, several techniques for prefetching have been investigated.
Especially pointer-based data structures which suffer from \textquotedbl pointer
chasing\textquotedbl , as for example large graphs \citep{ainsworth2016graph,ainsworth2017software}
and databases \citep{jonathan2018exploiting,psaropoulos2017interleaving,psaropoulos2019interleaving},
profit from prefetching. The main challenge is to issue the prefetch
instructions early enough such that the data is loaded timely into
cache, but also late enough such that it does not clog the cache.

Recent work has focused on introducing code stages to deal with dynamic
memory access and uncertainty in the number of lookups. This is achieved
either by manually implementing state machines \citep{kocberber2015asynchronous}
or usage of coroutines \citep{jonathan2018exploiting,psaropoulos2017interleaving,psaropoulos2019interleaving}.
Coroutines are resumable functions that can suspend their current
execution. In both cases a prefetching instruction is inserted just
before the memory stalls. After the prefetch instruction the program
saves its current state and continues with other work. When entering
the function the next time the required piece of memory is available
and operation can resume where it left off. These techniques are promising
for cases where irregularity such as variable-length pointer chains
and potential early loop exits prevent usage of simpler prefetching
techniques.

If the number of pointer dereferences is known ahead of time and is
constant across lookups, a promising technique to employ is group
prefetching \citep{chen2007improving}. Group prefetching is a loop-transformation
method which breaks a single for loop into an outer and several inner
loops allowing batchwise processing of code stages and critical data
to be prefetched.

Another technique capable of hiding cache misses is software pipelining
\citep{lam1988software,allan1995software,watanabe2019simd}. Here
loops are transformed such that the instructions inside of the loop
are carried out with an offset and overlapped with each other (details
in \prettyref{subsec:lagging-rb}). Thus, memory accesses and arithmetic
operations inside the loop are no longer mutually dependent. As recent
CPUs are superscalar, independent memory accesses and arithmetic operations
can be executed in parallel. This increases the number of instructions
completed per cycle. As the modification of the for loop increases
the number of instructions, the increase in instructions per cycle
has to be greater than the increase in number of instructions to improve
the overall performance.

The present study begins with a description of our setting in terms
of hardware and software as well as the profiling framework (\prettyref{sec:benchmarking-framework}).
We use the open-source community simulation code NEST (\prettyref{subsec:NEST})
to obtain performance data from a real-world application and as a
framework for reference implementations of the optimization techniques
discussed. Absolute performance data require a concrete neuronal network
model close to the ones used in production. The network model introduced
in \prettyref{subsec:network-model} is prototypical for a wide class
of models in neuroscience, it is scalable, and it has been used in
a number of previous studies. A scientific community code needs to
be developed and maintained over decades. Therefore it is important
that new algorithms do not improve the performance on one architecture
while making it impossible to adapt to the next generation of systems.
Therefore, we assess the performance of a recent mainstream architecture,
a common but older high-end cluster, and a dedicated supercomputer
(\prettyref{subsec:Systems}). \prettyref{sec:memory-access-spike-delivery}
describes the data structures representing neurons and synapses and
how spikes travel through these data structures from arrival at the
compute node to their ultimate delivery at the target neurons. Based
on this, the subsequent two sections form the core of the investigation.
\prettyref{subsec:latency-hiding} proposes new algorithms based on
latency-hiding techniques indicated above for different phases of
spike delivery. \prettyref{sec:results} presents a quantitative analysis
of the effects of the new algorithms and their combination on different
hardware architectures. Finally, \prettyref{sec:discussion} derives
a combination of algorithms delivering a robust overall performance
gain across problem sizes and hardware architectures. The study concludes
by setting the results into the context of the generic problem of
applications with essentially random memory access patterns and their
implications for the interpretation of the von Neumann bottleneck
and the design of neuromorphic computing systems.

Source code, simulation and analysis scripts are openly available
\citep{pronold_jari_routing_caching_code} at Zenodo\footnote{\url{https://www.zenodo.org}}.
The presented conceptual and algorithmic work is part of our long-term
collaborative project to provide the technology for neural systems
simulations \citep{Gewaltig_07_11204}. Preliminary results have been
presented in abstract form \citep{Kunkel19_ISC}.

\section{Benchmarking framework}

\label{sec:benchmarking-framework}

\subsection{Simulation engine}

\label{subsec:NEST}In the present work we evaluate the concepts
and new algorithms in the framework of the simulation code NEST\footnote{\url{https://www.nest-simulator.org}}
\citep{Gewaltig_07_11204}, a widely used engine for spiking neuronal
networks of natural density at the resolution of individual nerve
cells and synapses. NEST is an open source project governed by the
public society NEST Initiative\footnote{\url{https://www.nest-initiative.org}}
and a component of the ICT infrastructure\footnote{\url{https://www.ebrains.eu}}
created by the European Human Brain Project (HBP). The development
is managed via GitHub where contributions undergo a formal code review
and consistency is ensured by continuous integration using automated
style checks and testing. The kernel of the simulator is written in
C++ and uses $MT$ coequal OpenMP \citep{OpenMPSpec} threads for
parallelization which are arranged into $M$ MPI \citep{MPIForum09}
processes harboring $T$ threads each. The differential equations
and state transitions defining neuron and synapse models are expressed
in the domain specific language NESTML \citep{Plotnikov16_93,linssen_charl_2020_3697733}
which generates the required C++ code for dynamic loading into the
simulation engine. The configuration of simulation experiments, including
neuron and synapse models, network structure and recorded data, are
specified interactively via Python using PyNEST \citep{Eppler09_12,Zaytsev14_23}.
Neurons are distributed across parallel resources together with their
incoming synapses in a round-robin fashion. This distribution implements
a simple load balancing scheme as it minimizes the number of neurons
from the same population, which may exhibit similar activity patterns,
on the same thread. Already \citet{Morrison05a} find that not only
the propagation of the dynamical state of the neuronal network but
also network construction needs to be parallelized to achieve sufficient
performance for practical applications. More than a decade later \citet{Ippen2017_30}
improve the algorithms for multi-threading on many-core systems and
point out that non-blocking memory allocation is essential for performance.
However, as network creation is not in the focus, the present work
stays with the system \texttt{malloc()}. The work is based on commit
059fe89 of release 2.18.

NEST uses a globally time-driven simulation scheme \citep{Morrison05a},
where neurons are typically updated every $0.1\ms$ and spike times
are constrained to this time grid. There is a biophysical delay between
the emission of a spike by the source neuron and the arrival at the
target neuron. Therefore, it suffices to exchange spike data between
threads in intervals of the minimal delay in the neuronal network
\citep{Morrison08_267}. Consequently, the simulation cycle propagating
the dynamical state of the network divides into three phases: update
neurons, communicate spikes between threads, and deliver spikes to
target neurons including the propagation of synaptic dynamics (see
\prettyref{fig:conclusions-simtime-contributions}). Different neuron
models require solvers of different computational load ranging from
precalculated exact propagator matrices for linear neuron models \citep{Rotter99a}
to generic solvers for non-linear differential equations with adaptive
time-stepping. Similarly the workload of synapses depends on the chosen
model. Static synapses are stateless whereas for plastic synapses
the state may depend on the activities of the pre- and the postsynaptic
neuron \citep{Morrison08_459}. Over the years the scalability of
NEST has been demonstrated on a range of supercomputers \citep{Helias12_26,Kunkel2012_5_35,Kunkel14_78,Jordan18_2}.
Recent revisions of the code employ MPI\_Alltoall to send spikes only
to MPI ranks where they have targets \citep{Jordan18_2}.

\subsection{Network model}

\label{subsec:network-model}To measure and compare the suggested
algorithmic improvements we use a balanced random network model \citep{Brunel00_183}
as a benchmark similar to the one used in previous studies on neuronal
network simulation technology \citep{Morrison07_1437,Helias12_26,Kunkel2012_5_35,Kunkel14_78,Ippen2017_30,Kunkel2017_11,Jordan18_2}.
The parameters for this benchmark model are specified in the parameter
tables 1, 2 and 3 in \citet{Jordan18_2}. Note that in contrast to
previous studies, where excitatory-excitatory connections exhibited
spike-timing dependent plasticity, the model considered here uses
static synapses, which do not exhibit any dynamics and consequently
have a fixed weight.

We consider the model a scalable version of a typical neuronal network
simulation as the neuronal activity exhibits an asynchronous irregular
spike pattern and does not depend significantly on the model's size.
Furthermore, the random connectivity of the network model represents
a worst-case scenario in terms of network structure: Local connection
patterns cannot, not even in principle, be exploited by representing
subnetworks on a subset of available nodes as a single neuron connects
with equal probability with any other neuron in the network.

All measurements of runtime in this study refer to the actual simulation
time, or in short ``sim time'', where the network state is propagated;
measurements of network-construction time and initialization time
are not part of this study.

\subsection{Systems}

\label{subsec:Systems}The detailed specifications of the computer
systems are given in \citep{pronold2022}. JURECA CM \citep{Krause:850758}
consists of $1872$ compute nodes (dual Intel Xeon E5-2680 v3 Haswell
12-core CPUs at $2.5\GHz$), DEEP-EST\footnote{\url{https://www.deep-projects.eu}}
has $50$ nodes (dual Intel Xeon Gold $6146$ Skylake 12-core CPUs
at $3.2\GHz$), and the K computer \citep{Miyazaki12} houses $82,944$
nodes ($8$-core Fujitsu SPARC64 VIIIfx CPU at $2\GHz$).

\subsection{Measurements of runtime and profiling}

\label{subsec:JUBE-NESTtimers-VTune}For systematic benchmarking
we rely on the Jülich Benchmarking Environment (JUBE) \footnote{\url{https://www.fz-juelich.de/jsc/jube}}\citep{Luehrs16_432},
which is a software suite actively developed by the Jülich Supercomputing
Centre. For measuring the time consumption of different parts of the
code we rely on the  \texttt{Stopwatch} class distributed with NEST.
This class acts as a wrapper around\texttt{ gettimeofday()} which
is part of the header \texttt{<sys/time.h>} of the C POSIX library\footnote{\url{https://pubs.opengroup.org/onlinepubs/9699919799/idx/head.html}}.

We use the Microarchitecture Exploration analysis mode of the Intel
VTune Profiler\footnote{\url{https://software.intel.com/vtune}},
which provides detailed information on hardware usage. By specifying
the option \texttt{uarch-exploration}, the running process is periodically
interrupted enabling the sampling of hardware events from the processor.
These events are used for calculating predefined ratios, which are
reported once the program has finished. The study collects data only
for the first $64$ MPI processes and restricts measurements to the
spike-delivery phase. Specifically, we focus on the Clockticks per
Instructions Retired (CPI) event ratio. The CPI measure is calculated
via dividing the number of unhalted processor cycles (clockticks)
by the number of instructions retired. CPI indicates to what extent
latency affects the application's execution, with smaller values corresponding
to smaller latencies.

\section{Memory access during spike delivery}

\label{sec:memory-access-spike-delivery}In spiking neuronal network
simulation code the temporally sparse event-based communication between
neurons presents a challenging performance bottleneck for modern architectures
optimized for dense data but hence also an optimization opportunity.
Due to the connection data structures and the spike-delivery algorithm
of our reference implementation NEST (\prettyref{subsec:NEST}), delivery
of spikes to their targets involves frequent access to essentially
random memory locations. Therefore, memory access is difficult to
predict automatically, leading to long data access times due to ineffective
use of caches. The following two sections describe the connection
data structures and the spike-delivery algorithm that serves as a
starting point for the latency-hiding techniques investigated in this
study (\prettyref{subsec:latency-hiding}) and as a reference in the
quantitative analysis (\prettyref{sec:results}). This reference algorithm
is a result of the preparatory refactoring of the original data structures
and algorithm presented in \citet{pronold2022}, which already achieves
a significant reduction in simulation time (\prettyref{fig:results-REF}).

\subsection{Memory layout of synapses and neurons}

\label{subsec:fundamental-data-structures}
\begin{figure}
\begin{centering}
\includegraphics[width=1\columnwidth]{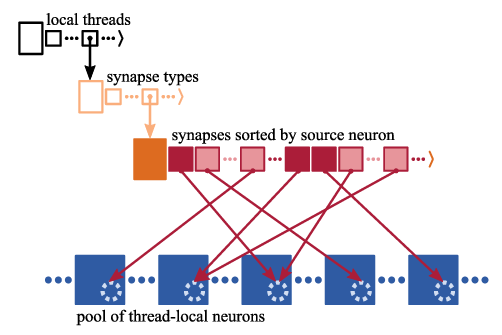}
\par\end{centering}
\caption{Memory layout of synapses and neurons on each MPI process. Each process
stores the local synapses (pink filled squares) in a three-dimensional
resizable array sorted by hosting thread and synapse type. At the
innermost level, synapses are arranged in source-specific target segments
(dark pink: first synapse; light pink: subsequent targets); only one
innermost array shown for simplicity. Target neurons (blue filled
squares) are stored in neuron-type and thread-specific memory pools;
only one pool shown for simplicity. Each neuron maintains a spike
ring buffer (dotted light blue circles). Synapses have access to their
target neurons' spike ring buffers through pointers (dark pink arrows).
Adapted from Figure 1 of \citet{pronold2022}.\label{fig:5g-data-structures}}
\end{figure}
Each synapse is represented on the same MPI process and thread as
its target neuron \citep{Morrison05a}, where a model synapse has
a memory footprint of few tens of Bytes, while a model neuron easily
consumes a Kilobyte or more. Each MPI process makes use of a three-dimensional
resizable array to store the process-local synapses sorted by hosting
thread and synapse type (\prettyref{fig:5g-data-structures}). The
data structure takes into account that each neuron typically connects
to many target neurons (out-degree): In the innermost arrays, synapses
are sorted by source neuron and thereby arranged in target segments,
each consisting of at least one target synapse potentially followed
by subsequent targets. The number of unique source neurons and hence
the average length of target segments depends on the distribution
of synapses across MPI processes and threads as well as the number
of synapse types. In the limit of sparsity where the number of neurons
in the network exceeds the number of thread-local synapses, the length
of target segments approaches one.

In order to account for synaptic transmission delays each neuron maintains
a ring buffer accommodating the incoming spikes until they are due
to take effect on the neuronal dynamics \citep{Morrison05a}. In the
reference algorithm, a synapse has access to its target neuron's spike
ring buffer through a pointer \citep[cf. Section 4.2,][]{pronold2022}.

Neurons of the same type that are also hosted by the same thread belong
to the same memory pool consisting of multiple chunks that allow for
contiguous storage of many objects. As synapses from many different
source neurons converge on a target neuron (in-degree), the memory
locations of target neurons cannot be ordered according to source
neurons and are hence independent of the order of synapses in the
target segments.

\subsection{Spike-delivery algorithm}

\label{subsec:reference-deliver-algo}
\begin{algorithm}[t]
\SetAlgorithmName{}{}{}
\SetAlgoRefName{bwRB*}
\DontPrintSemicolon
\SetKwFunction{Send}{Send}
\SetKwFunction{AddValue}{AddValue}
\SetKw{In}{in}
\SetKw{To}{to}
\SetKw{True}{true}
\SetKw{BRB}{\normalfont B\_RB}
\KwData{$spike\_reg$, $synapses$}
\BlankLine
create arrays $target\_rb$, $delay$, $weight$ of size \BRB\;
$i \leftarrow 0$\;
\ForEach{$spike$ \In $spike\_reg$}{
  $lcid \leftarrow spike.lcid$\;
  $subsq \leftarrow \True$\;
  \nlset{TS}\While{$subsq$}{
    \nlset{SYN}$(subsq,\,target\_rb[i],\,delay[i],\,weight[i]) \leftarrow synapses[lcid].$\Send{}\;
    $i \leftarrow i+1$\;
    $lcid \leftarrow lcid+1$\;
    \If{$i==$ \BRB}{
      \For{$i \leftarrow 0$ \To $\BRB-1$}{
        \nlset{RB*}prefetch $target\_rb[i]$\;
      }
      \For{$i \leftarrow 0$ \To $\BRB-1$}{
        \nlset{RB}$target\_rb[i].$\AddValue{$delay[i]$, $weight[i]$}\;
      }
      $i \leftarrow 0$\;
    }
  }
}
process remaining entries in $target\_rb$\;

\caption{Delivery of spikes in batches of size $B_{\mathrm{RB}}$, including
group prefetching of spike ring buffers. \textbf{\footnotesize{}TS}
marks iteration over a synaptic target segment.\textbf{ }\textbf{\footnotesize{}SYN}
marks access to an individual target synapse;\textbf{\footnotesize{}
RB} marks access to the spike ring buffer of the corresponding target
neuron;\textbf{\footnotesize{} RB{*}} marks group prefetching prior
to access. Based on \prettyref{alg:REF}.\label{alg:bwRB*}}
\end{algorithm}
The distributed time-driven simulation of spiking neuronal networks
proceeds in a cycle of updates to all neurons, communication of all
recent spikes across MPI processes, and delivery of the spikes to
their process-local synaptic and neuronal targets. After every spike
communication, each MPI process holds a receive buffer filled with
spike data that need to be dispatched to the process-local targets.
Each spike entry addresses an entire target segment of synapses (\prettyref{subsec:fundamental-data-structures}).
To this end, the spike entry needs to be able to locate the beginning
of the target segment within the three-dimensional data structure
storing the process-local synapses (\prettyref{fig:5g-data-structures}),
i.e. the first synapse of the target segment. Therefore, each spike
entry conveys identifiers for the hosting thread and the type of the
synapse, as well as the synapse's index within the innermost resizable
array.

In the reference algorithm, the delivery of spikes from the MPI receive
buffer to the local targets hosted by different threads is a two-step
process allowing for an entirely thread-parallel delivery with a single
synchronization point \citep[cf. Section 4.1,][]{pronold2022}. First,
the threads sort the spike entries by hosting thread and synapse type
in parallel using a dedicated intermediate data structure, called
spike-receive register, only then the threads dispatch the spikes,
now exclusively reading relevant entries.

Starting with the first synaptic target, the hosting thread subsequently
processes all synapses of a spike's target segment. The number of
spike entries in the receive buffer depends on the degree of distribution
of synapses across MPI processes and threads as well as the number
of synapse types. Arranging synapses in source-specific target segments
such that only one spike needs to be communicated to address an entire
segment is an effective optimization for small to medium-scale simulations
(see Section 3.3 in \citealp{Jordan18_2}). However, when increasing
the network size in a weak-scaling experiment, ever more source neurons
have ever fewer thread-local targets, or conversely, ever more thread-local
synapses originate from different source neurons such that they can
no longer be combined into long target segments addressed by a single
source-specific spike entry. This is the limit of sparsity discussed
in \prettyref{subsec:fundamental-data-structures}.

The synapse object stores all information relevant for the subsequent
delivery to the target neuron, foremost a pointer to the target neuron's
spike ring buffer. During delivery of the spike to the target neuron
the algorithm retrieves the pointer from the synapse as well as synaptic
properties such as delay and weight, which define time and amplitude
of the spike's impact on the neuron, respectively. Taking into account
the delay, the algorithm adds the weight of the incoming spike to
the correct position in the neuronal spike ring buffer.

\begin{algorithm}[t]
\SetAlgorithmName{}{}{}
\SetAlgoRefName{REF}
\DontPrintSemicolon
\SetKwFunction{Send}{Send}
\SetKwFunction{AddValue}{AddValue}
\SetKw{In}{in}
\SetKw{True}{true}
\KwData{$spike\_reg$, $synapses$}
\BlankLine
\ForEach{$spike$ \In $spike\_reg$}{
  $lcid \leftarrow spike.lcid$\;
  $subsq \leftarrow \True$\;
  \nlset{TS}\While{$subsq$}{
    \nlset{SYN}$(subsq,\,target\_rb,\,delay,\,weight) \leftarrow synapses[lcid].$\Send{}\;
    $lcid \leftarrow lcid+1$\;
    \nlset{RB}$target\_rb.$\AddValue{$delay$, $weight$}\;
  }
}

\caption{Reference algorithm delivering spikes to local targets. \textbf{\footnotesize{}TS}
marks iteration over a synaptic target segment.\textbf{ }\textbf{\footnotesize{}SYN}
marks access to an individual target synapse; \textbf{\footnotesize{}RB}
marks access to the spike ring buffer of the corresponding target
neuron. Variables typeset in italics, functions in typewriter. See
\citet{pronold2022} for a more detailed presentation.\label{alg:REF}}
\end{algorithm}
For each relevant spike entry the hosting thread accesses target synapses
and neurons in an alternating fashion. All target synapses of a spike
entry are in contiguous locations in memory as they are part of the
same target segment, but the corresponding target neurons are in nonadjacent
memory locations. Furthermore, with every spike entry, the thread
proceeds to a different synaptic target segment, most likely not in
a proximate memory location. With increasing sparsity of the network
in a weak-scaling experiment, such switches between target segments
become more frequent as ever more spike entries need to be delivered
to ever shorter target segments. In the sparse limit, both accessing
target synapses and the corresponding target neurons requires the
hosting thread to jump to random memory locations. The pseudocode
of the reference algorithm (\prettyref{alg:REF}) presents this memory
bottleneck of the spike-delivery algorithm in an abstract way omitting
intricacies caused by support for multi-threading and different synapse
types.

\section{Latency-hiding techniques}

\label{subsec:latency-hiding}

Based on the reference algorithm (\prettyref{alg:REF}), we investigate
three techniques to hide memory fetch latency and reduce the number
of cache misses during spike delivery. To assist the reader in following
the algorithmic changes, we provide pseudocode for the reference algorithm
(\prettyref{alg:REF}), each of the potential adaptations (\prettyref{alg:bwRB*},
\prettyref{alg:lagRB}, and \prettyref{alg:bwTS}), and a combination
of two adaptations (\prettyref{alg:bwTSRB*}). 
\begin{algorithm}[h]
\SetAlgorithmName{}{}{}
\SetAlgoRefName{lagRB}
\DontPrintSemicolon
\SetKwFunction{Send}{Send}
\SetKwFunction{AddValue}{AddValue}
\SetKw{In}{in}
\SetKw{True}{true}
\SetKw{False}{false}
\SetKw{And}{and}
\SetKw{BRB}{\normalfont B\_RB}
\KwData{$spike\_reg$, $synapses$}
\BlankLine
create arrays $target\_rb$, $delay$, $weight$ of size \BRB+1\;
$i \leftarrow 0$\;
$j \leftarrow 0$\;
\ForEach{$spike$ \In $spike\_reg$}{
  $lcid \leftarrow spike.lcid$\;
  $is\_init \leftarrow \True$\;
  $subsq \leftarrow \True$\;
  \nlset{TS}\While{$subsq$}{
    \nlset{SYN}$(subsq,\,target\_rb[i],\,delay[i],\,weight[i]) \leftarrow synapses[lcid].$\Send{}\;
    $i \leftarrow i+1$\;
    \If{$i==\BRB+1$}{
      $i \leftarrow 0$\;
    }
    $lcid \leftarrow lcid+1$\;
    \If{$is\_init$ \And $i==$ \BRB}{
      $is\_init \leftarrow$ \False\;
    }
    \Else{
      \nlset{RB}$target\_rb[j].$\AddValue{$delay[j]$, $weight[j]$}\;
      $j \leftarrow j+1$\;
      \If{$j==\BRB+1$}{
        $j \leftarrow 0$\;
      }
    }
  }
}
process remaining entries in $target\_rb$\;

\caption{Delivery of spikes introducing a lagged access to ring buffers with
respect to the corresponding target synapses; lag is $B_{\mathrm{RB}}$.
\textbf{\footnotesize{}TS} marks iteration over a synaptic target
segment.\textbf{ }\textbf{\footnotesize{}SYN} marks access to an individual
target synapse;\textbf{\footnotesize{} RB} marks access to the ring
buffer of the target neuron\textbf{. }Based on \ref{alg:REF}.\label{alg:lagRB}}
\end{algorithm}
The algorithms \prettyref{alg:bwRB*} and \prettyref{alg:lagRB} encode
competing adaptations of the reference algorithm \prettyref{alg:REF},
whereas \prettyref{alg:bwTS} encodes an adaptation that can be combined
with either of the former. The pseudocode is reduced to the essential
elements of the spike delivery; in particular, we apply the following
simplifications: the code does not take into account multiple threads
or synapse types and the initial thread-parallel transfer of spike
entries from the MPI receive buffer to the spike-receive register
is omitted (\prettyref{subsec:reference-deliver-algo}). The performance
of the algorithms is analyzed in \prettyref{sec:results}.

The reference algorithm has access to a resizable array of thread-local
$synapses$ and to a spike register ($spike\_reg$), which contains
all spike entries that need to be delivered \citep[see Section 4.1 of][for details]{pronold2022}.
For each spike entry, the location of the first target synapse is
extracted and assigned to the variable $lcid$, which is then used
in the enclosed while loop to iterate over the spike's entire synaptic
target segment within the $synapses$ array (\textbf{\footnotesize{}TS}).
Each synapse stores an indicator ($subsq$) of whether the target
segment continues or not. To deliver a spike to the target synapse
at position $lcid$, the synapse member function \texttt{Send()} is
called on $synapses[lcid]$ returning the indicator $subsq$, the
pointer to the spike ring buffer, and the synaptic delay and weight
(\textbf{\footnotesize{}SYN}). The pointer is then used to call \texttt{AddValue()},
a member function of the neuronal spike ring buffer requiring the
delay and the weight (\textbf{\footnotesize{}RB}). Taking into account
the delay, the spike ring buffer adds the weight of the incoming spike
to the correct position. This implements the delivery of the spike
to the target neuron. Note the dependency of the algorithmic step
\textbf{\footnotesize{}RB} on step \textbf{\footnotesize{}SYN}, which
is readily visible as a result of the prior refactoring \citep[Section 4.2 of][]{pronold2022}.

\subsection{Batchwise access to spike ring buffers}

\label{subsec:batchwise-rb}We next consider a loop-transformation
method that allows for group prefetching \citep{chen2007improving}.
The method breaks a single for loop of $L$ iterations containing
several code stages into an outer and several inner loops of size
$L_{B}$ and $B$, respectively, where $L=L_{B}B$. For example, an
original loop $\left\{ X_{i}Y_{i}\right\} ^{L}$ over two operations
$X_{i}$ and $Y_{i}$ depending on running index $i$ is transformed
into $\left\{ X_{(j-1)B+1}\ldots X_{jB}Y_{(j-1)B+1}\ldots Y_{jB}\right\} ^{L_{B}}$.
The code stages of the original for loop are hence processed in a
batchwise manner. All code stages are handled $B$ times before moving
to the next step of the outer loop. By introducing code stages of
size $B$, a prior loop prefetching all critical data can be inserted,
making use of memory-level parallelism.

We adapt the reference algorithm such that access to synapses and
access to spike ring buffers is carried out in batches of size $B_{\mathrm{RB}}$
(\prettyref{alg:bwRB*}). First, $B_{\mathrm{RB}}$ synapses are accessed
to retrieve the pointers to the corresponding target neurons' spike
ring buffers as well as the synaptic weights and delays (\textbf{\footnotesize{}SYN}).
The retrieved synaptic information is temporarily buffered in three
auxiliary arrays ($\ensuremath{target\_rb}$, $delay$, $weight$)
of size $B_{\mathrm{RB}}$. Each time $B_{\mathrm{RB}}$ synapses
have been accessed such that the auxiliary buffers are filled, batchwise
access to the $B_{\mathrm{RB}}$ collected target spike ring buffers
is triggered to add the corresponding weights to the correct buffer
positions (\textbf{\footnotesize{}RB}). An optional prior step is
the software-induced prefetching of the $B_{\mathrm{RB}}$ spike ring
buffers into the cache (\textbf{\footnotesize{}RB{*}}). To this end,\texttt{
}in the NEST implementation (\prettyref{subsec:NEST}) we make use
of the GCC built-in function \texttt{\_\_builtin\_prefetch(const void
{*}addr, ...)}\footnote{\url{https://gcc.gnu.org/onlinedocs/gcc/Other-Builtins.html}};
the prefetching can be enabled at compile time. The method is referred
to as group prefetching as it consecutively prefetches data from an
entire batch of $B_{\mathrm{RB}}$ individual memory addresses before
operating on the group of data instead of using a per-memory-address
approach alternating between prefetching and processing. In the quantitative
analysis (\prettyref{sec:results}) we refer to this set of optimizations
as either bwRB or \prettyref{alg:bwRB*}, where the asterisk indicates
prefetching.

The batchwise progression is agnostic with regard to the boundaries
of synaptic target segments (\prettyref{subsec:fundamental-data-structures}),
which means that in case of short target segments, it can take several
iterations of the for loop over spike entries to process $B_{\mathrm{RB}}$
synaptic targets. Leaving and re-entering of the enclosed while loop
traversing the synaptic target segment of a specific spike entry (\textbf{\footnotesize{}TS})
does not affect the progression. Eventually, all spike entries in
the register are processed and either delivered to their synaptic
targets or added to the auxiliary array. After the loop over the spike
register exits the algorithm delivers the remaining entries in the
auxiliary arrays to their targets.

\subsection{Lagging access to spike ring buffers}

\label{subsec:lagging-rb}This optimization exploits the idea of
software pipelining for spike delivery. In software pipelining \citep{lam1988software,allan1995software,watanabe2019simd}
loops are reformed in such a way that the instructions inside of the
loop are carried out with an offset of $B\geq1$ and overlapped with
each other. For example, an original loop of $L$ iterations $\left\{ X_{i}Y_{i}\right\} ^{L}$
over two operations $X_{i}$ and $Y_{i}$ depending on index $i$
is transformed into $\left\{ X_{i}\right\} ^{B}\left\{ X_{i}Y_{i-B}\right\} _{B+1}^{L}\left\{ Y\right\} _{L-B+1}^{L}$.
By doing so, the operation $X$ inside of the central loop is from
a different iteration than $Y$.

We adapt the reference algorithm such that access to synapses and
access to spike ring buffers is still carried out in an alternating
fashion but algorithmic progression is always $B_{\mathrm{RB}}$ synapses
ahead of spike ring buffers (\prettyref{alg:lagRB}). This means when
a ring buffer is accessed the synaptic information and the respective
pointers for the next $B_{\mathrm{RB}}$ spike ring buffers are already
available in local arrays. To this end, $B_{\mathrm{RB}}$ synapses
are initially accessed to retrieve the pointers to the corresponding
target neurons' spike ring buffers as well as the synaptic weights
and delays (\textbf{\footnotesize{}SYN}). The retrieved synaptic information
is temporarily buffered in auxiliary arrays ($\ensuremath{target\_rb}$,
$delay$, $weight$) of size $B_{\mathrm{RB}}+1$. Once $B_{\mathrm{RB}}$
synapses have been accessed such that the auxiliary buffers are almost
full (\texttt{if}-condition is met), the initialization phase ends
and alternating access to synaptic targets (\textbf{\footnotesize{}SYN})
and neuronal spike ring buffers (\textbf{\footnotesize{}RB}, \texttt{else}-clause)
starts, using the auxiliary arrays as ring buffers and the indices
$i$ and $j$ for writing to the arrays and reading from the arrays,
respectively. In the quantitative analysis (\prettyref{sec:results})
we refer to this optimization as \prettyref{alg:lagRB}.

As for \prettyref{alg:bwRB*}, leaving and re-entering of the enclosed
while loop traversing the synaptic target segment of a specific spike
entry (\textbf{\footnotesize{}TS}) does not interrupt the cyclic processing
of the auxiliary arrays. Eventually, all spike entries in the register
are processed and delivered to their synaptic targets, but the auxiliary
arrays still contain synaptic information that needs to be delivered
to the spike ring buffers.

\subsection{Batchwise access to target segments}

\label{subsec:batch-wise-target-segments}
\begin{algorithm}[t]
\SetAlgorithmName{}{}{}
\SetAlgoRefName{bwTS}
\DontPrintSemicolon
\SetKwFor{RepeatTimes}{repeat}{times}{}
\SetKwFunction{Send}{Send}
\SetKwFunction{AddValue}{AddValue}
\SetKwFunction{Size}{Size}
\SetKwFunction{GetTSSize}{GetTSSize}
\SetKw{In}{in}
\SetKw{To}{to}
\SetKw{Next}{\normalfont next}
\SetKw{Of}{\normalfont of}
\SetKw{True}{true}
\SetKw{BTS}{\normalfont B\_TS}
\KwData{$spike\_reg$, $synapses$}
\BlankLine
create arrays $lcid$, $ts\_size$ of size \BTS\;
$l \leftarrow 0$\;
\RepeatTimes{$spike\_reg.$\Size{} $/$ \BTS}{
  \For{$k \leftarrow 0$ \To $\BTS-1$}{
    $lcid[k] \leftarrow spike\_reg[l+k].lcid$\;
  }
  $l \leftarrow l+\BTS$\;
  \For{$k \leftarrow 0$ \To $\BTS-1$}{
    $ts\_size[k] \leftarrow synapses[lcid[k]].$\GetTSSize{}\;
  }
  \For{$k \leftarrow 0$ \To $\BTS-1$}{
    \nlset{TS}\RepeatTimes{$ts\_size[k]$}{
      \nlset{SYN}$(target\_rb,\,delay,\,weight) \leftarrow synapses[lcid[k]].$\Send{}\;
      $lcid[k] \leftarrow lcid[k]+1$\;
      \nlset{RB}$target\_rb.$\AddValue{$delay$, $weight$}\;
    }
  }
}
process remaining entries in $spike\_reg$\;

\caption{Delivery of spikes with processing of spike entries and corresponding
target segments in batches of size $B_{\mathrm{TS}}$. \textbf{\footnotesize{}TS}
marks iteration over a synaptic target segment using a fixed count
loop according to the target-segment size $ts\_size$.\textbf{ }\textbf{\footnotesize{}SYN}
marks access to a target synapse;\textbf{\footnotesize{} RB} marks
access to the spike ring buffer of the corresponding target neuron.
Based on \prettyref{alg:REF}.\label{alg:bwTS}}
\end{algorithm}
\begin{algorithm}[!t]
\SetAlgorithmName{}{}{}
\SetAlgoRefName{bwTSRB*}
\DontPrintSemicolon
\SetKwFor{RepeatTimes}{repeat}{times}{}
\SetKwFunction{Send}{Send}
\SetKwFunction{AddValue}{AddValue}
\SetKwFunction{Size}{Size}
\SetKwFunction{GetTSSize}{GetTSSize}
\SetKw{In}{in}
\SetKw{To}{to}
\SetKw{Next}{\normalfont next}
\SetKw{Of}{\normalfont of}
\SetKw{True}{true}
\SetKw{BTS}{\normalfont B\_TS}
\SetKw{BRB}{\normalfont B\_RB}
\KwData{$spike\_reg$, $synapses$}
\BlankLine
create arrays $lcid$, $ts\_size$ of size \BTS\;
create arrays $target\_rb$, $delay$, $weight$ of size \BRB\;
$l \leftarrow 0$\;
$i \leftarrow 0$\;
\RepeatTimes{$spike\_reg.$\Size{} $/$ \BTS}{
  \For{$k \leftarrow 0$ \To $\BTS-1$}{
    $lcid[k] \leftarrow spike\_reg[l+k].lcid$\;
  }
  $l \leftarrow l+\BTS$\;
  \For{$k \leftarrow 0$ \To $\BTS-1$}{
    $ts\_size[k] \leftarrow synapses[lcid[k]].$\GetTSSize{}\;
  }
  \For{$k \leftarrow 0$ \To $\BTS-1$}{
    \nlset{TS}\RepeatTimes{$ts\_size[k]$}{
     \nlset{SYN}$(target\_rb[i],\,delay[i],\,weight[i]) \leftarrow synapses[lcid[k]].$\Send{}\;
      $i \leftarrow i+1$\;
      $lcid[k] \leftarrow lcid[k]+1$\;
      \If{$i==$ \BRB}{
        \For{$i \leftarrow 0$ \To $\BRB-1$}{
          \nlset{RB*}prefetch $target\_rb[i]$\;
        }
        \For{$i \leftarrow 0$ \To $\BRB-1$}{
          \nlset{RB}$target\_rb[i].$\AddValue{$delay[i]$, $weight[i]$}\;
        }
        $i \leftarrow 0$\;
      }
    }
  }
}
process remaining entries in $spike\_reg$ and $target\_rb$\;\;

\caption{Delivery of spikes combining processing of spike entries and corresponding
target segments in batches of size $B_{\mathrm{TS}}$ (\prettyref{alg:bwTS})
with access to target synapses and neuronal ring buffers in batches
of size $B_{\mathrm{RB}}$, including group prefetching of spike ring
buffers (\prettyref{alg:bwRB*}). \textbf{\footnotesize{}TS} marks
iteration over a synaptic target segment using a fixed count loop
according to the target-segment size $ts\_size$.\textbf{ }\textbf{\footnotesize{}SYN}
marks access to an individual target synapse;\textbf{\footnotesize{}
RB} marks access to the ring buffer of the corresponding target neuron;\textbf{\footnotesize{}
RB{*}} marks group prefetching prior to access. Based on \prettyref{alg:REF}.\label{alg:bwTSRB*}}
\end{algorithm}
The previous two algorithms, \prettyref{alg:bwRB*} and \prettyref{alg:lagRB},
are concerned with disentangling access to synapses and neuronal spike
ring buffers. They implicitly make hopping from one synaptic target
segment to the next more seamless as they progress writing and reading
the auxiliary arrays ($\ensuremath{target\_rb}$, $delay$, $weight$)
irrespective of the boundaries of single target segments. The optimizations
of the present algorithm, \prettyref{alg:bwTS}, directly target the
iteration over spike entries in the spike receive register ($spike\_reg$)
in the outermost loop and the concomitant processing of synaptic target
segments of different lengths in different memory locations. The quantitative
analysis (\prettyref{sec:results}) refers to this set of optimizations
as \prettyref{alg:bwTS}.

We adapt the reference algorithm to process spike entries in the spike
receive register in batches of size $B_{\mathrm{TS}}$, which in turn
entails batchwise access to $B_{\mathrm{TS}}$ distinct target segments
(\prettyref{alg:bwTS}). Like in \prettyref{alg:bwRB*}, this optimization
is based on a loop-transformation method (\prettyref{subsec:batchwise-rb})
but targets a different part of the algorithm. Moreover, we replace
the while loop iterating over every target synapse of a specific target
segment with a fixed count loop (\textbf{\footnotesize{}TS}). This
requires however that the length of the target segment is available
when entering the fixed count loop. Here, we decide on a straightforward
solution: the information is provided by the first target synapse
of each target segment. To this end, we extend synapse objects with
a member variable to store the target-segment size, which needs to
be determined just once when all synapses have been created (not shown
in the algorithm); the public member function \texttt{GetTSSize()}
returns the size. Note that while the algorithm only requires this
capability for the first synapse of each target segment, all synapse
objects are equipped with the extra member variable as they are stored
in a container for homogeneous objects (\prettyref{subsec:fundamental-data-structures}).
In our reference implementation (\prettyref{subsec:NEST}), we ensure
that the extra member variable does not increase the per-synapse memory
usage by reducing the storage size of another synaptic member variable,
namely the synaptic delay. This does not affect the precision of results
for our benchmark network model as the reduced storage size is sufficient
to fully represent the homogeneous and relatively short delays of
the model (\prettyref{subsec:network-model}). This solution does,
however, not generalize to all types of network models as, for example,
in case of longer delays, an increase in per-synapse memory usage
due to the additional member variable might be inevitable.

For each batch of $B_{\mathrm{TS}}$ spike entries, the algorithm
carries out three consecutive for loops each with a fixed number of
$B_{\mathrm{TS}}$ iterations. In the first for loop iterating over
$B_{\mathrm{TS}}$ spike entries, the locations of the first target
synapses of the corresponding target segments are extracted from the
spike entries and buffered in the auxiliary array $lcid$. In the
subsequent for loop the $B_{\mathrm{TS}}$ locations are used to access
the first target synapses to retrieve the corresponding target-segment
sizes, which are buffered in the auxiliary array $ts\_size$. In the
final for loop and the enclosed fixed count loop, alternating access
to synaptic targets (\textbf{\footnotesize{}SYN}) and neuronal spike
ring buffers (\textbf{\footnotesize{}RB}) is carried out like in the
reference algorithm. If the number of spike entries in the spike receive
register is not divisible by $B_{\mathrm{TS}}$ without remainder,
the remaining spike entries are processed in a similar fashion using
three consecutive for loops, where the fixed number of iterations
is given by the number of remaining entries.

\subsection{Combined batchwise access to target segments and spike ring buffers}

\label{app:appendix-bwTSRB}The combined algorithm \prettyref{alg:bwTSRB*}
adopts the loop structure of \prettyref{alg:bwTS}, but the instructions
inside the fixed count loop iterating over a specific synaptic target
segment (\textbf{\footnotesize{}TS}) are adapted according to \prettyref{alg:bwRB*}.
The combined algorithm requires all auxiliary arrays of the two individual
algorithms, where indices $l$ and $k$ and index $i$ are used as
in \prettyref{alg:bwTS} and \prettyref{alg:bwRB*}, respectively.

\section{Results}

\label{sec:results}
\begin{figure}[t]
\begin{centering}
\includegraphics{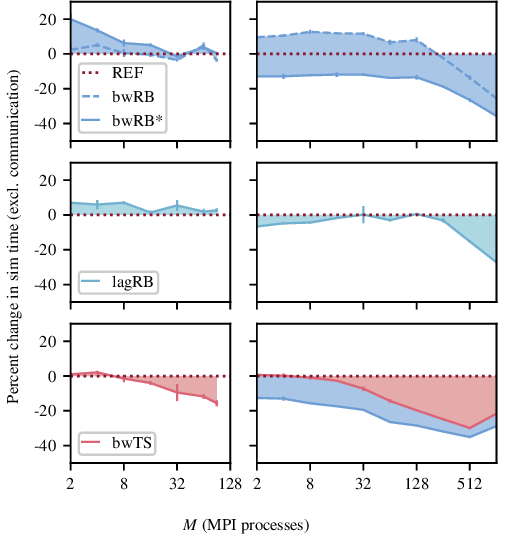}
\par\end{centering}
\caption{Cumulative change in simulation time relative to refactored code disregarding
communication as a function of number of MPI processes. Left column
DEEP-EST CM and right column JURECA CM: linear-log representation
for number of MPI processes $M\in\{2;\:4;\:8;\:16;\:32;\:64;\:90\}$
and $M\in\{2;\:4;\:8;\:16;\:32;\:64;\:128;\:256;\:512;\:1,024\}$,
respectively. Dark carmine red dotted line at zero percent (\prettyref{alg:REF},
\prettyref{subsec:reference-deliver-algo}) indicates performance
of reference code. First row, batchwise access to spike ring buffers
(\prettyref{subsec:batchwise-rb}) with batches of size $16$ without
group prefetching (bwRB, dashed light blue curve) and with prefetching
(\prettyref{alg:bwRB*}, solid light blue curve). Second row, software
pipelining with a lag of $16$ (\prettyref{alg:lagRB}, turquoise
curve). Third row, batchwise access to target segments with batches
of size $16$ (\prettyref{alg:bwTS}, solid coral red curve), for
JURECA CM (right) further combined with batchwise access to spike
ring buffers using prefetching (\prettyref{alg:bwRB*}, same coloring
as in top row: solid light blue curve; see \prettyref{alg:bwTSRB*}
for combined pseudocode). Shadings fill area to respective reference.
Weak scaling of benchmark network model as in \prettyref{fig:conclusions-simtime-contributions}.\label{fig:SEA-SWP-NPP125000}}
\end{figure}
We quantitatively evaluate the effect of the three different techniques
of latency hiding \prettyref{alg:bwRB*}, \prettyref{alg:lagRB},
and \prettyref{alg:bwTS} on simulation time relative to the refactored
code (\prettyref{subsec:reference-deliver-algo}), where the optimization
\prettyref{alg:bwTS} can be combined with either \prettyref{alg:bwRB*}
or \prettyref{alg:lagRB} because they modify different parts of the
code. As the combined optimizations may either support each other
or in the worst case reduce the impact of the best individual optimization,
we also present performance data for a combined scenario of \prettyref{alg:bwTS}
and \prettyref{alg:bwRB*}. \prettyref{fig:SEA-SWP-NPP125000} summarizes
the quantitative results. The section concludes by demonstrating that
the improved performance is indeed a result of a reduction of clock
ticks per instruction retired (\prettyref{subsec:results-vtune}).

Plain batchwise access to spike ring buffers without group prefetching
(bwRB, \prettyref{subsec:batchwise-rb}) has no effect on the performance
on the DEEP-EST CM and enabling prefetching (\prettyref{alg:bwRB*})
actually worsens the situation for small numbers of MPI processes.
The situation is entirely different on JURECA CM which uses an older
generation of processors. Here, plain batchwise access has a negative
effect on the performance but group prefetching leads to an overall
performance improvement. At larger numbers of MPI processes already
the batchwise processing increasingly improves performance and the
additional gain by group prefetching remains constant. For tested
batch sizes between $1$ and $64$, we observe the least decline in
performance for batch sizes of $8$ or larger on DEEP-EST CM and the
best performance for batch sizes of $16$ or larger on JURECA CM (data
not shown). However, a comprehensive analysis is outside the scope
of this study; we do not claim that this observation generalizes to
other architectures.

An alternative latency-hiding technique for the same part of the code
is lagged access to ring buffers (\prettyref{alg:lagRB}, \prettyref{subsec:lagging-rb}),
where we use a lag of $16$. The algorithm thus retrieves information
from a target synapse $16$ steps ahead from the position in the target
segment where it currently accesses the corresponding ring buffer,
thereby decoupling these operations. Again we observe almost no effect
on the DEEP-EST CM. JURECA CM exhibits an increasing gain only for
large numbers of MPI processes. For tested lags between $1$ and
$16$, we observe little differences on DEEP-EST CM and the best performance
for lags of $2$ and larger on JURECA CM (data not shown).

The introduction of the spike-receive register \citep{pronold2022}
opens the opportunity for batchwise processing of the register and
the synaptic target segments (\prettyref{alg:bwTS}, \prettyref{subsec:batch-wise-target-segments}).
The batchwise processing of target segments has an increasing gain
reaching $20\%$ at $90$ MPI processes on the DEEP-EST CM, and the
total gain reaches more than $40\%$ on JURECA CM. For tested batch
sizes between $1$ and $64$, we observe little differences on DEEP-EST
CM and the best performance for batch sizes of up to $16$ on JURECA
CM (data not shown).

Finally, on JURECA CM we investigate a combined implementation of
batchwise access to target segments and batchwise access to spike
ring buffers including group prefetching (\prettyref{alg:bwTSRB*})
as the two techniques modify different parts of the code. We do not
further consider lagged access to ring buffers (\prettyref{subsec:lagging-rb}),
as \prettyref{alg:lagRB} is only effective for large numbers of MPI
processes, while its alternative \prettyref{alg:bwRB*} exhibits a
sustained gain over the full range. The lower right panel of \prettyref{fig:SEA-SWP-NPP125000}
shows that \prettyref{alg:bwRB*} is effective already at small numbers
of MPI processes but continues to improve the performance of \prettyref{alg:bwTS},
leading to a combined gain of $50\%$ at larger numbers of MPI processes.

The preparatory refactoring described in \citet{pronold2022} and
the latency-hiding techniques investigated in this study jointly improve
simulation time with respect to the original code (\prettyref{fig:conclusions-simtime-contributions}).
While refactoring is most effective for small numbers of MPI processes
where synaptic target segments are long (\prettyref{fig:results-REF}),
latency hiding is most effective for larger numbers of MPI processes
where target segments significantly shorten. The differential behavior
of the combined technologies leads to a sustained performance gain
of $30\%$ to $50\%$ over the whole range of investigated MPI processes.

\subsection{Dependence of effectiveness on microarchitecure}

\label{subsec:results-vtune}

\begin{figure}
\includegraphics{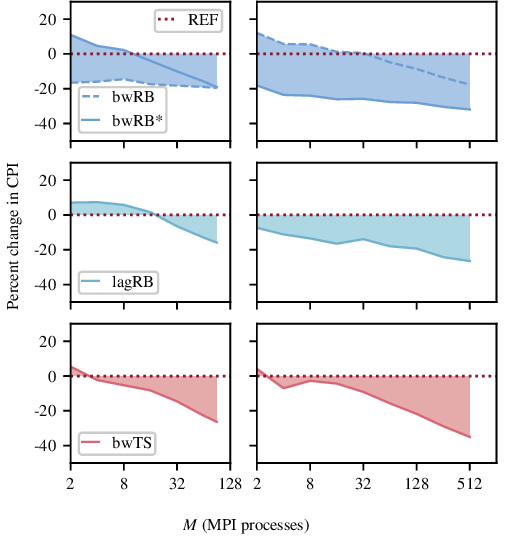}\caption{Relative change in clock ticks per instruction retired (CPI) during
spike delivery as a function of the number of MPI processes. Same
arrangement of panels and labeling as in \prettyref{fig:SEA-SWP-NPP125000}.
\label{fig:Vtune_metrics}}
\end{figure}
The latency-hiding techniques introduced in \prettyref{subsec:latency-hiding}
promise to speed up the overall code by reducing the latency of memory
access. Indeed our measurements show that our particular application
becomes substantially faster, despite additional lines of code required
to implement the techniques. Profiling tools (\prettyref{subsec:JUBE-NESTtimers-VTune})
provide direct access to the latency that processor instructions
experience. This allows us to investigate whether the new algorithms
are faster due to the reduction of latency. Specifically we are investigating
clock ticks per instruction retired (CPI), a metric available in the
tool VTune (\prettyref{subsec:JUBE-NESTtimers-VTune}). All algorithms
exhibit a decrease in CPI compared to \prettyref{alg:REF} (\prettyref{fig:Vtune_metrics}),
except for less than $32$ MPI processes \prettyref{alg:bwRB*} on
DEEP-EST CM and bwRB for JURECA CM. Larger networks show larger gains
and \prettyref{alg:lagRB} and \prettyref{alg:bwTS} show similar
behavior on both machines. The dependence on network size is particularly
pronounced for \prettyref{alg:bwTS}. This directly relates to the
decreasing length of target segments, diminishing the algorithmic
potential for ordering synaptic data for sequential processing. As
\prettyref{alg:bwTS} addresses the bottleneck of hopping from one
target segment to the next, its effect increases as target-segment
lengths decrease. The characteristics of bwRB (\prettyref{subsec:batchwise-rb})
and \prettyref{alg:bwRB*} on DEEP-EST and JURECA appear inverted.
This hints that the processor in DEEP-EST CM, in contrast to JURECA
CM, automatically prefetches the necessary data on time. Adding additional
explicit prefetching instructions may interfere with the automatic
prefetching and hence degrade performance. While the changes in simulation
time (\prettyref{fig:SEA-SWP-NPP125000}) follow the changes in CPI,
the reduction in sim time does not fully reflect the success observed
in CPI. All latency-hiding techniques come with additional lines of
source code. This may lead to an increased number of processor instructions
that the reduction in CPI has to make up for, but the compiler may
now also find more opportunities for reducing the number of instructions.

\section{Discussion}

\label{sec:discussion}

At the outset of this investigation stands the observation that
for a typical neuronal network the spike-delivery phase dominates
total simulation time in state-of-the-art simulation code and that
the time required for spike delivery increases under weak scaling
such that it still dominates at $1024$ MPI processes (\prettyref{fig:conclusions-simtime-contributions}).
This article explores techniques to re\-arrange the elementary algorithmic
steps required to deliver the incoming essentially random spike data
to the thread-local targets such that they can be more efficiently
processed by conventional computer hardware. Each of the techniques
under consideration is characterized individually, but some address
different steps of spike delivery. Combining the most promising techniques
achieves a significant performance boost.

The combined spike-delivery algorithm starts with the thread-parallel
sorting of the spike data from the MPI receive buffer into a thread-specific
data structure called spike receive register (SRR), where sorting
is performed according to hosting thread and synapse type of each
spike's synaptic targets. This preparatory refactoring improves parallelization
but is not related to the optimization of cache performance on the
level of the individual threads. The technical companion paper \citep{pronold2022}
provides the detailed description of the SRR together with pseudo
code. Sorting into the spike receive register is followed by the thread-parallel
processing of the SRR in fixed-size batches cycling through three
stages (\prettyref{alg:bwTS}, \prettyref{subsec:batch-wise-target-segments}):
for each batch of spike entries, collecting first the location and
second the length of the spike's synaptic target segment in separate
arrays before continuing with the actual delivery as a third stage.
In the third stage, also synapses and corresponding neuronal spike
ring buffers are processed in batches (\prettyref{alg:bwRB*}, \prettyref{subsec:batchwise-rb})
by, irrespective of target-segment boundaries, weight, delay, and
pointer to ring buffer, first collecting separate arrays for each
batch of synapses before adding the weights to the buffers.

As a result of the redesign the time required for spike delivery
is halved (reduced by factor of $2.1$) for two MPI processes compared
to the original algorithm and reduced to a third (by a factor of $2.9$)
for $1024$ MPI processes. In the original algorithm, the increased
network sparsity at larger numbers of MPI processes causes an increase
in spike-delivery time by a factor of $4.5$ between two and $1024$
MPI processes. The new algorithm reduces this to a factor of $3.3$.
As a consequence, simulation time is reduced by $43\%$ for low numbers
of MPI processes, by $49\%$ for $512$ MPI processes, where communication
is not yet dominating, and by $28\%$ for $1024$ MPI processes, where
communication time starts to dominate overall simulation time. The
measurements were obtained using ParaStationMPI v5.2.2-1, while initial
tests using the more recent version v5.4 result in a significant reduction
in communication time (data not shown). However, following up on this
observation is not within the scope of this manuscript due to JURECA
CM being decommissioned. In summary, with the new algorithm spike
delivery still dominates the simulation time below $1024$ MPI processes
under weak scaling, but it increases less rapidly such that now even
below $1024$ MPI processes the major cause for the loss in performance
is the increase in communication time. A second qualitative change
sets in at $1024$ MPI processes, where the absolute simulation time
is no longer dominated by spike delivery but by communication time.
Hence, the new algorithm overcomes the previously reported barrier
of spike delivery limiting the performance on supercomputers (Figure
$12$ in \citealp{Jordan18_2}). Now progress can be made by optimizing
communication, for example, by exploiting the spatial organization
of neuronal networks. Cortical neuronal networks are characterized
by a local coupling with a space constant of a few hundred micrometers
and delays in the range of a tenth of a millisecond combined with
long distance coupling between brain areas and delays beyond a millisecond.
If cortical areas were represented on one or a few compute nodes,
the communication between nodes hosting different areas could be reduced
to much larger intervals than required between nodes hosting neurons
of the same area. Topology-aware distribution of neuronal networks
across compute nodes has been exploited in other simulation codes
\citep{Kozloski11_5_15,Igarashi2019_13_71} but without taking into
account different delay categories.

Performance profiling of the algorithms \prettyref{alg:bwRB*}, \prettyref{alg:lagRB},
and \prettyref{alg:bwTS} indicates that they indeed reduce latency
(\prettyref{subsec:results-vtune}). In accordance with the measurements
of sim times (\prettyref{sec:results}), which show that the techniques
are more successful on JURECA CM (Intel Haswell) than on DEEP-EST
CM (Intel Skylake), the relative reduction in the performance metric
clock ticks per instruction retired (CPI) is also more pronounced
on JURECA CM than on DEEP-EST CM. We speculate that the improved cache
utilization of the newer generation processor (Skylake) renders \prettyref{alg:bwRB*}
and \prettyref{alg:lagRB} ineffective or even detrimental, at least
in the regime up to $90$ MPI processes investigated here. On JURECA
CM the effect on sim time increases beyond $128$ MPI processes, which
is beyond the current capacity of DEEP-EST CM. Therefore, emulating
simulations on more than $128$ MPI processes on DEEP-EST CM using
the NEST dry-run mode \citep{Kunkel2017_11} may provide further insights.
A comprehensive comparison between the two generations of processors
based on microbenchmarks as presented in \citep{alappat2020} for
the related microarchitectures Intel Broadwell and Intel Cascade Lake
is out of the scope of this study. It is our hope though that our
efforts to present the changes to code in an abstract fashion make
this study relevant for a broader computer science community and might
even inspire the definition of future microbenchmarks. The present
work focuses on the level of source code only. It may be illuminating
to explore in future studies the effect of algorithmic changes on
the level of the resulting assembler code.

The incoming spike events of a compute node specify the thread hosting
the target neuron as well as the location of the synaptic targets,
but are unsorted with respect to the hosting thread and synapse type.
Nevertheless the present work shows that the processing of spikes
can be completely parallelized requiring only a single synchronization
between the threads at the point where the spikes have been sorted
according to target threads and synapse types, which is when all spikes
have been transferred from the MPI receive buffer into the novel spike
receive register. This suggests that spike delivery can fully profit
from a further increase in number of threads per compute node. Due
to the large number of outgoing synapses per neuron, early work on
distributed simulations of spiking neuronal networks employing tens
of compute nodes was concerned with the problem of efficiently delivering
each spike emitted by a specific source neuron on a specific MPI process
to many targets on all MPI processes \citep{Morrison05a}. With increasing
parallelization using more compute nodes and threads, however, each
neuron has ever fewer thread-local targets resulting in ever shorter
synaptic target segments in the data structure storing the local synapses.
Therefore, later work concentrated on reducing the memory overhead
caused by the increasing number of shorter target segments \citep{Kunkel14_78}.
However, shorter target segments are also a burden computationally
because memory locations need to be switched often leaving little
opportunity for vectorization. The present study overcomes the problem
of short segments by reorganizing the spike delivery algorithm such
that it operates across the boundaries of target segments and in this
way becomes more independent of the degree of parallelization. Spikes
are routed to their thread-local targets in a batchwise fashion using
the technique of loop transformation. This explicit declaration of
virtually independent code blocks apparently helps compilers to generate
efficient machine code.

Due to the simplicity of the neuron model of the benchmark and the
lack of synaptic plasticity, the application has little workload in
terms of the propagation of neuron and synapse dynamics and thus exposes
bottlenecks in the delivery of spikes to local targets and the communication
between compute nodes. The synaptic delay of about a millisecond assumed
in the present benchmark model is used in prominent neuronal network
models of the balanced random class \citep{Brunel00_183} and representative
for the connectivity at the brain scale \citep{Schmidt18_1409}. In
models of the local network below a square millimeter of cortical
surface \citep[cf.][]{Potjans14_785} the minimal delay (\prettyref{subsec:NEST})
is an order of magnitude smaller than the delay considered in this
study. Consequently, the simulation of such models requires a ten-fold
increase in the number of communication calls and therefore the communication
phase takes up a larger fraction of the total simulation time. Nevertheless,
for a given spike rate the amount of spikes that need to be delivered
in a given stretch of biological time is independent of the number
of communication calls. Therefore, we expect the optimizations discussed
in the present work to still be effective, while their impact on total
simulation time will be lower. Neuroscientists have started to investigate
models combining the local structure of the brain with its organization
over long distances \citep{Schmidt18_e1006359}. The minimal delay
in such a brain-scale network is the minimal delay in the local structure.
Unless the placement of neurons on compute nodes respects the architecture
of the network, global communication in intervals as for the local
network is required, limiting the success of optimizations of spike
delivery.

Our analysis is restricted to static neuronal networks. In such networks
all synaptic connections and their weights are determined at the time
of network construction. In nature synaptic efficacies change over
time: the phenomenon called synaptic plasticity. A further mechanism,
removing and creating synapses, is called structural plasticity. A
wide class of models of synaptic plasticity can be formulated as an
update scheme driven by the presynaptic spike events in which the
synaptic weight is only computed for those times where it becomes
visible for the postsynaptic neuron \citep{Morrison07_1437}. In this
way the computations become part of the spike delivery phase and contribute
substantially to its duration. Depending on the particular plasticity
model, alongside the location of the ring buffer, additional information
on the target neuron needs to be accessed. This may be the time of
the last spike, the full record of spikes since the last presynaptic
spike, or the membrane potential \citep{Stapmanns21_609147}. Our
companion paper \citep{pronold2022} discusses the resulting tradeoff
between memory consumption and speed. A domain specific language like
NESTML \citep{Plotnikov16_93} could avoid cluttering the model pool
with solutions for different optimization goals by providing directives
influencing the balance between the generation of more compact or
faster code depending on the needs. How the more complex data handling
for plastic synapses and the floating point operations on the synaptic
weights interact with the batchwise processing needs to be evaluated.

The communication scheme between compute nodes assumed in the present
study sends for each spike a separate event to each thread containing
at least one target neuron. This enables a straightforward readout
of the MPI receive buffer containing the incoming spikes (\prettyref{subsec:reference-deliver-algo}).
However, for typical cortical connection densities where each neuron
has on the order of $10,000$ postsynaptic targets for simulations
with less than $10,000$ compute nodes and more than one thread per
node, compute nodes receive the same spike event multiple times. This
overhead may challenge the bandwidth of the communication network
and limit the speed of the communication phase. Future work needs
to investigate whether duplicate spike events can be efficiently
compressed such that the costs of decompression in the spike-delivery
phase are smaller than the gain in the communication phase.

The present study analyses spike-delivery times and overall simulation
times in a weak-scaling scenario. Neuroscientists and industry may,
however, also be interested in reducing the simulation time of an
application for a fixed size network studied over long stretches of
biological time as required for system-level learning or for the capability
to interact with the real world in robotics. Therefore strong-scaling
scenarios are of interest. With an increasing number of threads, the
number of target neurons and therefore synapses a thread needs to
take care of decreases. Consequently the amount of memory a thread
operates on decreases. This suggests that it becomes easier for the
core running the thread to predict memory access and if the amount
of cache memory per core decreases proportionally to the number of
cores faster execution is expected. In this way, the combination of
a highly parallel spike-delivery algorithm with a many-core architecture
overcomes the von Neumann bottleneck of conventional applications.
Thus our findings reconsidering the spike-delivery algorithm support
a more positive view on the prospects of improving the performance
of neuronal network simulations by prefetching and pipelining than
the analytical results of \citep{cremonesi2020analytic,cremonesi2020understanding}.
It remains to be seen whether the optimizations by batchwise processing
explored in the present work become less relevant in situations where
each core needs to manage a small amount of memory anyway.

The study removes several bottlenecks in routing spikes in a compute
node to ever more distributed targets. The spike-delivery phase of
a simulation code for spiking neuronal network remains a specific
sorting problem in space and time, but our study exposes that its
fully parallelizable nature will certainly benefit from future fine-grained
processing hardware.

\section*{Funding}

Partly supported by the European Union's Horizon 2020 (H2020) funding
framework under grant no. 785907 (Human Brain Project, HBP SGA2),
no. 945539 (HBP SGA3), and no. 754304 (DEEP-EST), as well as the Helmholtz
Association Initiative and Networking Fund project no. SO-092 (Advanced
Computing Architectures, ACA), as well as the Deutsche Forschungsgemeinschaft
(DFG, German Research Foundation) - 368482240/GRK2416. Use of the
JURECA supercomputer in J\"ulich was made possible through VSR computation
time grant JINB33. This research used resources of K computer at the
RIKEN Center for Computational Science (R-CCS). Supported by the project
Exploratory Challenge on Post-K Computer (Understanding the neural
mechanisms of thoughts and its applications to AI) of the Ministry
of Education, Culture, Sports, Science and Technology (MEXT).

\section*{Acknowledgments}

We are grateful to Mitsuhisa Sato for his guidance, which helped us
shape the project, to Johanna Senk and Dennis Terhorst for fruitful
discussions and joint efforts at the HPC Optimisation and Scaling
Workshop 2019 at the Jülich Supercomputing Centre, to Sebastian Lührs
for help with JUBE, to Laurent Montigny and Hans-Christian Hoppe for
insights on the Intel microarchitectures, to our colleagues in the
Simulation and Data Laboratory Neuroscience of the J\"ulich Supercomputing
Centre for continuous collaboration, and to the NEST development community
for the concepts and implementation of NEST (\href{http://www.nest-simulator.org}{http://www.nest-simulator.org}).

\bibliographystyle{elsarticle-num-names-custom}

\end{document}